\documentclass[prb,aps,twocolumn]{revtex4-1}
\usepackage{bm}
\usepackage{amsmath}
\usepackage{amssymb} 
\usepackage{epsfig}
\usepackage{graphicx}
\usepackage{color}
\usepackage{xcolor}
\usepackage{ulem}
\usepackage{cancel}
\usepackage{hyperref} 
\newcommand{\pderiv}[2]{\frac{\partial #1}{\partial #2}}
\renewcommand{\phi}{\varphi}
\renewcommand{\kappa}{\varkappa}

\renewcommand{\i}{\mathrm i}
\newcommand{\e}{\mathrm e}
\newcommand{\eps}{\varepsilon}

\begin{document}

\title{Second harmonic generation at the edge of a two-dimensional electron gas}

\author{M.\,V.\,Durnev}
\author{S.\,A.\,Tarasenko}

\affiliation{Ioffe Institute, 194021 St.\,Petersburg, Russia}

\begin{abstract}  

We show that driving a two-dimensional electron gas by an in-plane electric field oscillating at the frequency $\omega$ gives rise to an electric current at $2\omega$ flowing near the edge of the system. This current has both parallel and perpendicular to the edge components, which emit electromagnetic waves at $2\omega$ with different polarizations. We develop a microscopic theory of such an edge second harmonic generation and calculate the edge current at $2\omega$ in different regimes of electron transport and electric field screening. We also show that at high frequencies the spatial profile of the edge current contains oscillations caused by excitation of plasma waves.

\end{abstract}
 
\maketitle

\section{Introduction} 

Non-linear transport and optical phenomena in two-dimensional (2D) electron systems are at the core of modern research in solid-state physics~\cite{Hendry2010,Glazov2014,Durnev2019,Calafell2021}.
Of particular interest are the second-order effects comprising second harmonic generation (SHG)~\cite{Zhang2020,Dean2009,Glazov2011,Mikhailov2011,Wang2016,Zhang2019,Bykov2012,Golub2014,Wehling2015,Ho2020,Wang2015} and generation of dc current by ac electric field of radiation~\cite{Ivchenko_book,Tarasenko2011,Drexler2013,Budkin2016,Kheirabadi2018,Quereda2018,IvchenkoGanichev2018,Kiemle2020}. Such effects in the leading electro-dipole electron-photon interaction occur in structures with broken space inversion symmetry and, therefore, have been established as a sensitive tool to probe structural inhomogeneity, crystalline symmetry, the staking and twist of 2D crystal flakes, etc.~\cite{Dean2009, Bykov2013,Li2013,Mennel2019,Zhou2020,Stepanov2020}. 

In small-size samples, the translational and inversion symmetry is naturally broken at the edges, which gives rise to additional (edge-related) sources
of second-order nonlinearity. The corresponding photogalvanic currents flowing along the edges and controlled by the electromagnetic field polarization have been observed in single and bilayer graphene~\cite{Karch2011,Candussio2020,Candussio2021,Candussio2021b}. 
A kinetic theory of the edge photogalvanic effects has been developed for the intraband (Drude-like) optical transitions~\cite{Karch2011,Candussio2020,Durnev_pssb2021}, inter-Landau level transitions~\cite{Candussio2021},  interband one-photon~\cite{Durnev2021} and two-photon absorption~\cite{Candussio2021b} in
2D Dirac materials. Edge effects in SHG response have been observed in 2D layers of transition metal dichalcogenides in the spectral range of interband transitions and attributed to the local modification of atomic and electronic structures at the edges~\cite{Yin2014,Mishina2015}.  The edge SHG induced by non-linear transport of 2D electron gas at the edge remains unexplored so far. 

Here, we study SHG induced by high-frequency transport of 2D electrons at the edge of a semi-infinite sample. We show that driving the electrons back and forth by an in-plane ac electric field at the frequency $\omega$ gives rise to an electric response at $2\omega$. The current at the double frequency emerges near the edge in a narrow region determined by the dynamical screening of the electric field and the mean free path of electrons. The current at $2\omega$ has both parallel and perpendicular to the edge components which have specific dependences on the incident field polarization and emit 
the electromagnetic field at $2\omega$ with different polarizations. We develop a kinetic theory of such an edge SHG and calculate the current at 
$2 \omega$ in different regimes of electron transport and electric field screening. At $\omega \tau_1 >1$, where $\tau_1$ is the momentum relaxation time of electrons, the spatial profile of the current contains oscillations caused by excitation of plasma waves~\cite{Volkov1988, Mikhailov2005,Zabolotnykh2019,Zagorodnev2021}. 

\section{Second harmonic emission by edge currents}

\begin{figure}[htpb]
\includegraphics[width=0.49\textwidth]{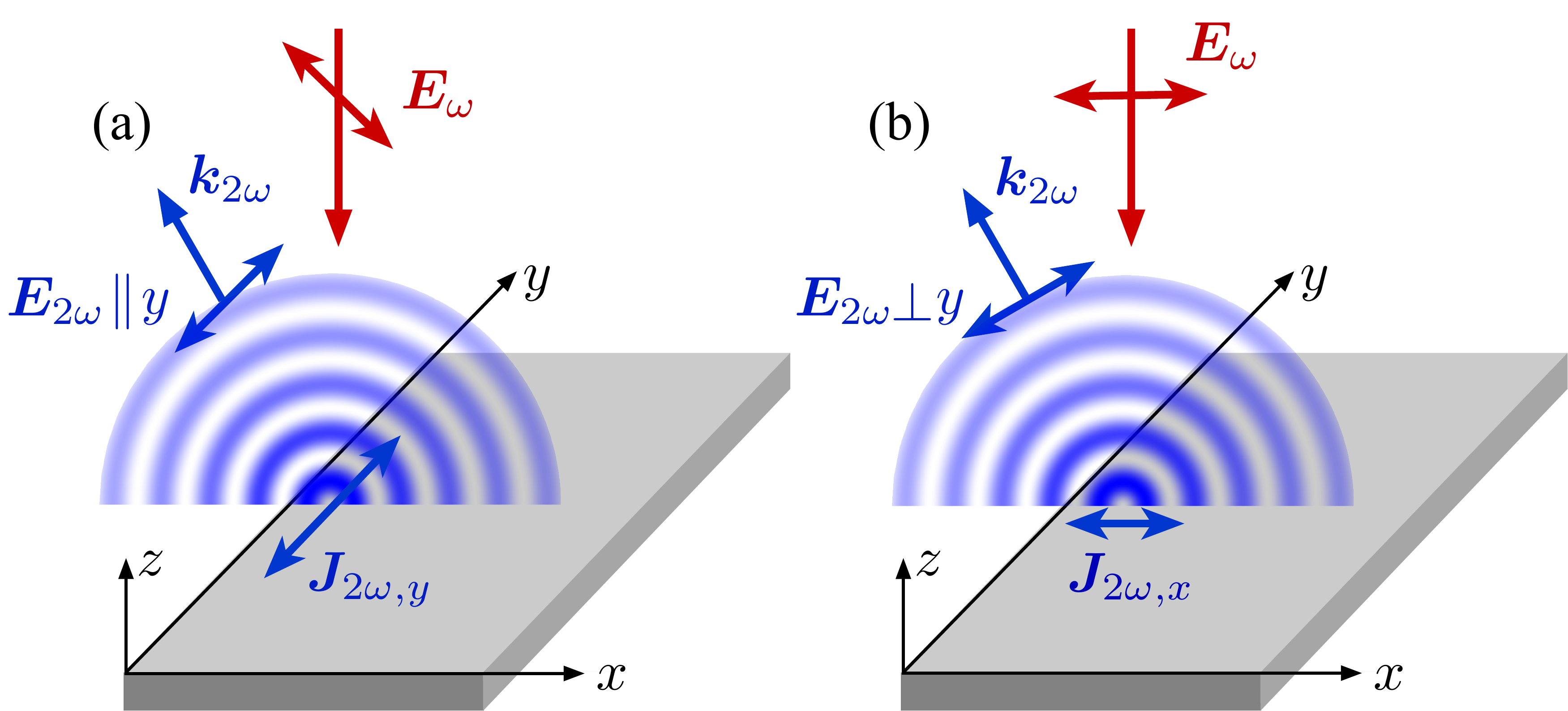}
\caption{\label{fig1} Second harmonic generation at the edge of 2DEG. Incident electromagnetic wave with the in-plane electric field $\bm E_\omega$ oscillating at the frequency $\omega$ induces the electric current $\bm J_{2\omega}$ at the double frequency
in a narrow strip near the edge. In turn, the edge current $\bm J_{2\omega}$ emits outgoing electromagnetic wave 
at $2\omega$ with the electric field amplitude $\bm E_{2\omega}$. 
(a) The incident field $\bm E_\omega$ with both $x$ and $y$ components induces the current along the edge $J_{2\omega, y} \propto E_{\omega, x} E_{\omega, y}$ emitting the wave with $\bm E_{2\omega} \parallel y$. (b) For the incident field $\bm E_{\omega} \parallel x$, the edge current flows perpendicular to the edge, $J_{2\omega, x} \propto E_{\omega, x}^2$, and the outgoing wave is polarized with  $\bm E_{2\omega} \perp y$.}
\end{figure}

Consider a semi-infinite 2DEG occupying the half-plane $x \geq 0$ at $z=0$ and irradiated by a plane electromagnetic wave with the electric field
$\bm E_\omega(t) = \bm E_\omega \e^{-\i\omega t} + \bm E_\omega^* \e^{\i\omega t}$, where $\bm E_\omega$ is the incident field amplitude,
see Fig.~\ref{fig1}. As we calculate below, non-linearity of the field-induced ac electron transport at the edge results in the emergence of the
current oscillating at the double frequency. This current has inhomogeneous density $\bm j_{2\omega} (x, t) = \bm j_{2\omega}(x) \e^{-2\i\omega t} + \bm j_{2\omega}^*(x) \e^{2\i\omega t}$ and is localized at the edge. The direction of the edge current depends on the incident field polarization. For the field polarized perpendicularly to the edge, $\bm E_\omega \perp y$ in Fig.~\ref{fig1}b, the current $\bm j_{2\omega}$ is perpendicular to the edge.
The field $\bm E_\omega$ with both parallel and perpendicular components induces also the current $\bm j_{2\omega}$ along the edge, Fig.~\ref{fig1}a.

The edge current, in turn, emits electromagnetic waves with the frequency $2\omega$ and the vector potential $\bm A_{2\omega}(\bm r, t) = \bm A_{2\omega}(\bm r) \e^{-2\i\omega t} + \bm A_{2\omega}^*(\bm r) \e^{2\i\omega t}$. The field $\bm A_{2\omega}(\bm r)$ can be found from the wave equation~\cite{Landau2}
\begin{equation} 
\label{wave}
\Delta \bm A_{2\omega} + k_{2\omega}^2 \bm A_{2\omega} = -\frac{4\pi}{c} \bm j_{2\omega}(x) \delta(z)\:,
%
\end{equation}
where  
$k_{2\omega} = 2\omega/c$ is the wave vector of the emitted wave and $\delta(z)$ is the Dirac delta-function.
Translation invariance in the $y$ direction implies that $\bm A_{2\omega}$ is independent of $y$.

The Green function of the two-dimensional Helmholtz equation enables us to present the solution of Eq.~\eqref{wave} 
in the form 
\begin{equation}
\label{A2w}
\bm A_{2\omega} (\bm r) = \frac{\i \pi}{c}  \int \bm j_{2\omega}(x') \mathrm{H}_0^{(1)} \left(k_{2\omega}\sqrt{(x-x')^2 + z^2}\right) dx'  ,
\end{equation}
where $\mathrm{H}_0^{(1)}$ is the Hankel function of the first kind. 

Far from the edge, $\bm A_{2\omega} (\bm r)$ represents the outgoing cylindrical wave. Its parameters can be found by analyzing the asymptotic of 
$\bm A_{2\omega} (\bm r)$ at large $R = \sqrt{x^2 + z^2}$. The asymptotic expansion of the Hankel function at large arguments has the form
$\mathrm{H}_0^{(1)}(\xi) \approx \sqrt{2/(\pi \xi}) \exp (\i \xi - \i\pi/4)$. Then, in the far field zone, i.e., at $R \gg 2\pi/k_{2\omega}$, and in the dipole approximation~\cite{Landau2}, which suggests $k_{2\omega} l \ll 2\pi$, where $l$ is the width of the stripe where the edge current flows, 
the field is given by
\begin{equation}  
\label{A2w_final}
\bm A_{2\omega} (\bm r) = \frac{\i \sqrt{2\pi}}{c \sqrt{k_{2\omega} R}} \exp\left[ \i \left( k_{2\omega} R - \frac{\pi}{4}\right) \right] \bm J_{2\omega}  \:,
\end{equation}
where
\begin{equation}
\label{J2w}
\bm J_{2\omega} = \int_0^{+\infty} \bm j_{2\omega}(x) dx
\end{equation}
is the total electric current at $2\omega$ flowing at the edge. 

The magnetic and electric fields of the outgoing wave are related to the vector potential as
\begin{equation}
\bm H_{2\omega} = \i \bm k_{2\omega}  \times \bm A_{2\omega} \:,~~~ \bm E_{2 \omega} = \bm H_{2\omega} \times \frac{\bm k_{2\omega}}{k_{2\omega}}\,,
\end{equation}
where $\bm k_{2\omega} = (x/R,0,z/R) k_{2\omega}$. The current $\bm J_{2\omega}$ flowing along the edge emits the electromagnetic waves with the field $\bm E_{2 \omega}$ parallel to the edge, whereas the edge current flowing perpendicularly to the edge induces the wave
with the field $\bm E_{2 \omega}$ lying in the $(x,z)$ plane, see Fig.~\ref{fig1}.

\section{Kinetic theory} 

Now we calculate the edge current $\bm J_{2\omega}$. In the kinetic approach, 
the response of an electron system to an external field is described by the Boltzmann equation for the electron distribution function $f$.
In our case $f=f(\bm p,x,t)$, and the Boltzmann equation has the form~\cite{Candussio2020,Durnev_pssb2021}
\begin{equation} 
\label{kinetic}
\pderiv{f}{t} + v_x \pderiv{f}{x} + e \bm{\mathcal E} \cdot \pderiv{f}{\bm p} = \mathrm{St}f \:,
\end{equation}
where $\bm p$ is the electron momentum, $\bm v = \bm p/m$ is the electron velocity, $m$ is the effective mass, $\bm{\mathcal E}(x,t)$ is the total electric field in the 2D layer acting on electrons, and $\mathrm{St}f$ is the collision integral. The collision integral describes the relaxation of electrons in the bulk of 
2D layer. Additionally, we assume that electrons are reflected specularly at the edge, which implies the boundary condition $f(p_x, p_y, x = 0) = f(-p_x, p_y, x = 0)$.

The field $\bm{\mathcal E}(x,t)$ is the sum of the incident field $\bm E_{\omega}(t)$ and the field induced by oscillating electric charge near the edge~\cite{Volkov1988}
%
\begin{equation} 
\label{screen}
\mathcal E_{x}(x,t) = E_{\omega,x}(t) +  \int\limits_0^{+\infty}\frac{2 \rho(x',t) dx'}{x-x'} \:,~~ \mathcal E_{y}(t) = E_{\omega,y}(t) \:, 
\end{equation}  
where $\rho(x,t)$ is the charge density given by $\rho(x,t) = e \nu \sum_{\bm p} (f - f_0)$, $\nu$ is the factor of spin and valley degeneracy, $f_0$ is the equilibrium distribution function, and the principal value of the integral in Eq.~\eqref{screen} is calculated. The charge density depends on the $x$ coordinate only, therefore the $y$ component of the electric field remains unscreened. Note, that we neglect electromagnetic retardation assuming that $\sigma_0 /{2\pi} \ll c$, where $\sigma_0$ is the two-dimensional conductivity of the electron gas~\cite{Volkov1988,Zagorodnev2021}. In fact, the same inequality justifies the dipole approximation used in Eq.~\eqref{A2w_final}. 

Since the external field is harmonic, we solve Eqs.~\eqref{kinetic} and~\eqref{screen} by expanding the distribution function $f$ and the electric field $\bm{\mathcal E}$ in the Fourier series as follows
\[
f(\bm p, x, t) = f_0 + [f_1(\bm p, x) \e^{-\i\omega t} + \mathrm{c.c.}] + [f_2(\bm p, x) \e^{-2\i\omega t} + \mathrm{c.c.}] \:,
\]
\begin{equation}
\bm{\mathcal E}(x,t) = [\bm{\mathcal E}_{\omega}(x) \e^{-\i\omega t} + \mathrm{c.c.}] + [\bm{\mathcal E}_{2\omega}(x) \e^{-2\i\omega t} + \mathrm{c.c.}]\:,
\end{equation}
where $f_1, \mathcal E_{\omega} \propto E_\omega$ and $f_2, \mathcal E_{2\omega} \propto E_{\omega}^2$ in the lowest order in the incident field amplitude. Note, that $f(\bm p, x, t)$ [as well as $\mathcal E_x(x,t)$] also contains time-independent non-equilibrium corrections $\propto E_{\omega}^2$. 
These corrections determine static edge polarization and dc edge currents~\cite{Candussio2020,Durnev_pssb2021}. However, they are not relevant for SHG and are omitted.
 
Equations for the corrections $f_1$ and $f_2$ read
\begin{align}
-\i \omega f_1 + v_x \pderiv{f_1}{x}  + e\bm{{\cal E}}_\omega \cdot \pderiv{f_0}{\bm p} = \mathrm{St} f_1   \:, \label{f1} \\
-2 \i \omega f_2 + v_x \pderiv{f_2}{x}  + e\bm{{\cal E}}_\omega \cdot \pderiv{f_1}{\bm p} + e \bm{{\cal E}}_{2\omega} \cdot \pderiv{f_0}{\bm p} = \mathrm{St} f_2 \:, \label{f2} 
\end{align} 
where 
\begin{equation}
\label{screen2}
\mathcal E_{n\omega, x}(x) = E_{\omega, x} \delta_{n,1}+ \int\limits_0^{+\infty}\frac{2 \rho_{n\omega}(x') dx'}{x-x'} \,,
\end{equation}
$\mathcal E_{n \omega, y} = E_{\omega, y} \delta_{n,1}$, and $\rho_{n \omega}(x) = e\nu \sum_{\bm p} f_{n}(\bm p,x)$.

The amplitude of the local current density oscillating at $2\omega$ is determined by the correction $f_2$ as follows
\begin{equation}
\bm{j}_{2\omega}(x) = e \nu \sum \limits_{\bm p} \bm v f_2(\bm p, x) 
\end{equation} 
and the total current is given by Eq.~\eqref{J2w}. 
Below we solve Eqs.~\eqref{f1}-\eqref{screen2} and calculate the current components parallel and perpendicular to the edge.

\section{Current along the edge}

Consider first the $y$ component of the edge current. Multiplying Eq.~\eqref{f2} by $v_y$ and summing up over $\bm p$ we obtain
\begin{align}
\sum_{\bm p} v_x v_y \pderiv{f_2}{x}
+ e \sum_{\bm p} v_y \left( \mathcal E_{\omega,x} \frac{\partial f_1}{\partial p_x} + E_{\omega,y} \frac{\partial f_1}{\partial p_y} \right) \nonumber \\
+ e \sum_{\bm p} v_y  \mathcal E_{2\omega,x} \frac{\partial f_0}{\partial p_x} =
(2 \i \omega - \tau_1^{-1} ) \sum_{\bm p} v_y f_2  \:,
\end{align}
where $\tau_1$ is the momentum relaxation time defined as $\sum_{\bm p} v_\alpha \mathrm{St} f = - \tau_1^{-1} \sum_{\bm p} v_\alpha f$.
Taking into account that $\sum_{\bm p} v_y \partial f_{n} /\partial p_x =0$ and $\sum_{\bm p} v_y \partial f_n /\partial p_y = - (1/m) \sum_{\bm p} f_n$
we obtain the current density
\begin{equation}
\label{jy}
j_{2\omega, y} = -\frac{e \nu \tau_1}{1 - 2\i\omega \tau_1} \left[ \sum \limits_{\bm p} v_x v_y \pderiv{f_2}{x} - \frac{e E_{\omega, y}}{m} \sum \limits_{\bm p} f_1 \right]  \:.
\end{equation} 

The total current $J_{2\omega, y}$ given by Eq.~\eqref{J2w} is found by integrating Eq.~\eqref{jy} over $x$. Using the relation $\sum_{\bm p} f_1 = -(\i/\omega) \sum_{\bm p} v_x \partial f_1/\partial x$, which follows from Eq.~\eqref{f1}, we obtain
\begin{multline}
\label{Jy}
J_{2\omega, y} = -\frac{e \nu \tau_1}{1 - 2\i\omega \tau_1} \sum \limits_{\bm p} v_x v_y \left[ f_2(\bm p, +\infty) - f_2(\bm p, 0) \right]  \\ -\frac{\i e^2 \nu \tau_1 E_{\omega, y}}{m \omega(1 - 2\i\omega \tau_1)} \sum \limits_{\bm p} v_x \left[  f_1(\bm p, +\infty) - f_1(\bm p, 0) \right] \:.
\end{multline}
Equation~\eqref{Jy} is general and does not rely on particular type of boundary conditions. It shows that the current at $2\omega$ emerges if the field-induced corrections to the electron distribution at the edge and the 2D bulk are different.

To  proceed further, we note that $\sum_{\bm p} v_x f_{1}(\bm p, 0) = 0$ since the current through the edge does not flow.
The sum $\sum_{\bm p} v_x v_y f_2(\bm p, 0)$ also vanishes for the specular reflection of electrons from the edge.
Therefore, the edge current $J_{2\omega, y}$ is determined by the corrections to the distribution function far from the edge, where the electric field is unscreened, i.e. $\bm{\mathcal E}_\omega = \bm E_\omega$ and $\bm{\mathcal E}_{2\omega} = 0$, and the electron distribution is homogeneous.

The term $\sum_{\bm p} v_x f_1(\bm p, +\infty)$ describes ac electric current and can be expressed via the bulk conductivity as follows
\begin{equation}
\label{jbulk}
e \nu \sum_{\bm p} \bm v f_1(\bm p, +\infty) = \sigma_\omega \bm E_\omega \:,
\end{equation}
where $\sigma_\omega = \sigma_0 / (1 - \i \omega \tau_1)$ is the conductivity at frequency $\omega$, 
$ \sigma_0 = n_e e^2 \tau_1/m$ is the static conductivity, and $n_e$ is the carrier density. 
The term $\sum_{\bm p} v_x v_y f_2(\bm p, +\infty)$  is calculated by multiplying Eq.~\eqref{f2} by $v_x v_y$ and summing up the result 
over $\bm p$, which gives
\begin{multline}
\label{vxvyf2}
\sum \limits_{\bm p} v_x v_y f_2(\bm p, +\infty) =  \frac{e \tau_2}{m(1 - 2\i\omega \tau_2)}  \\
\times \sum_{\bm p} f_1 (\bm p, +\infty) \left(v_x E_{\omega, y} + v_y E_{\omega, x} \right) ,
\end{multline}
where $\tau_2$ is the relaxation time of the second angular harmonic, $1/\tau_2 = - \sum_{\bm p} v_x v_y \mathrm{St} f/ \sum_{\bm p} v_x v_y f$.

Finally, taking into account Eqs.~\eqref{Jy}, \eqref{jbulk}, and \eqref{vxvyf2}, we obtain the current along the edge
%
\begin{equation}
\label{J2wy_final}
J_{2\omega, y} = - \frac{\i e \sigma_0 \tau_1(1 - 4\i\omega \tau_2)}{m\omega (1 - \i\omega \tau_1) (1 - 2\i\omega \tau_1) (1 - 2\i\omega \tau_2)} E_{\omega, x} E_{\omega, y}\:.
\end{equation}
The current $J_{2\omega, y}$ is proportional to $E_{\omega, x} E_{\omega, y}$. It reaches maxima for the field $\bm E_{\omega}$ linearly polarized at the angle $\pm \pi/4$ with respect to the edge and for circularly polarized field. The current vanishes for the field $\bm E_{\omega}$ polarized along or perpendicular to the edge.

\begin{figure}[htpb] 
\includegraphics[width=0.49\textwidth]{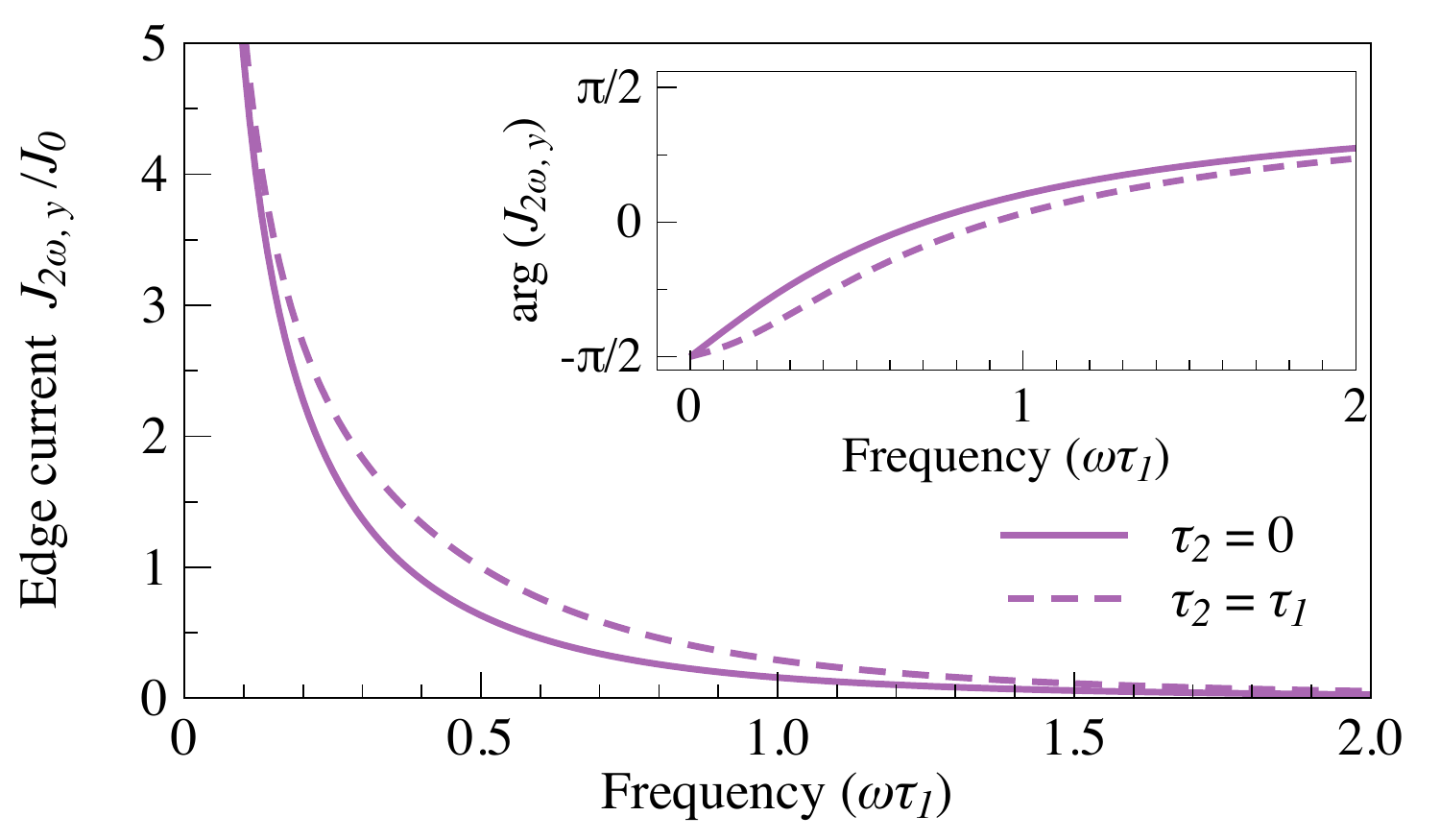}
\caption{\label{fig2} Frequency dependence of the edge current at $2\omega$ flowing parallel to the edge.  
The dashed line corresponds to short-range scattering with $\tau_1 = \tau_2$, the solid line stands for the hydrodynamic regime with $\tau_2 \ll \tau_1$. 
 The main graph and the inset show the modulus $|J_{2\omega, y}|$ and the argument $\mathrm{arg}(J_{2\omega, y})$ of the complex-value current $J_{2\omega, y}$, respectively. The current is measured in $J_0 = 2 e \sigma_0 \tau_1^2 E_{\omega,x} E_{\omega,y}/m$. } 
\end{figure}

 Figure~\ref{fig2} shows the frequency dependence of the current $J_{2\omega, y}$. The current is calculated after  Eq.~\eqref{J2wy_final} for linearly polarized incident field $\bm E_{\omega}$ and different ratio between the relaxation times $\tau_2$ and $\tau_1$. 
Since $J_{2\omega, y}$ is complex, both the modulus and the argument of $J_{2\omega, y}$ are plotted. 
We consider two cases: $\tau_2 = \tau_1$, which corresponds to electron scattering by short-range impurities, and $\tau_2 \ll \tau_1$, which corresponds to the hydrodynamic regime when frequent electron-electron collisions destroy the second angular harmonic of the distribution function. In the latter case, the first contribution to the edge current in Eq.~\eqref{Jy} vanishes. Figure~\ref{fig2} shows that $J_{2\omega, y}$ is weakly sensitive to the $\tau_2/\tau_1$ ratio. The current $\propto 1/\omega$ and $\propto 1/\omega^3$ at low and high frequencies, respectively.

\begin{figure}[htpb]
\includegraphics[width=0.49\textwidth]{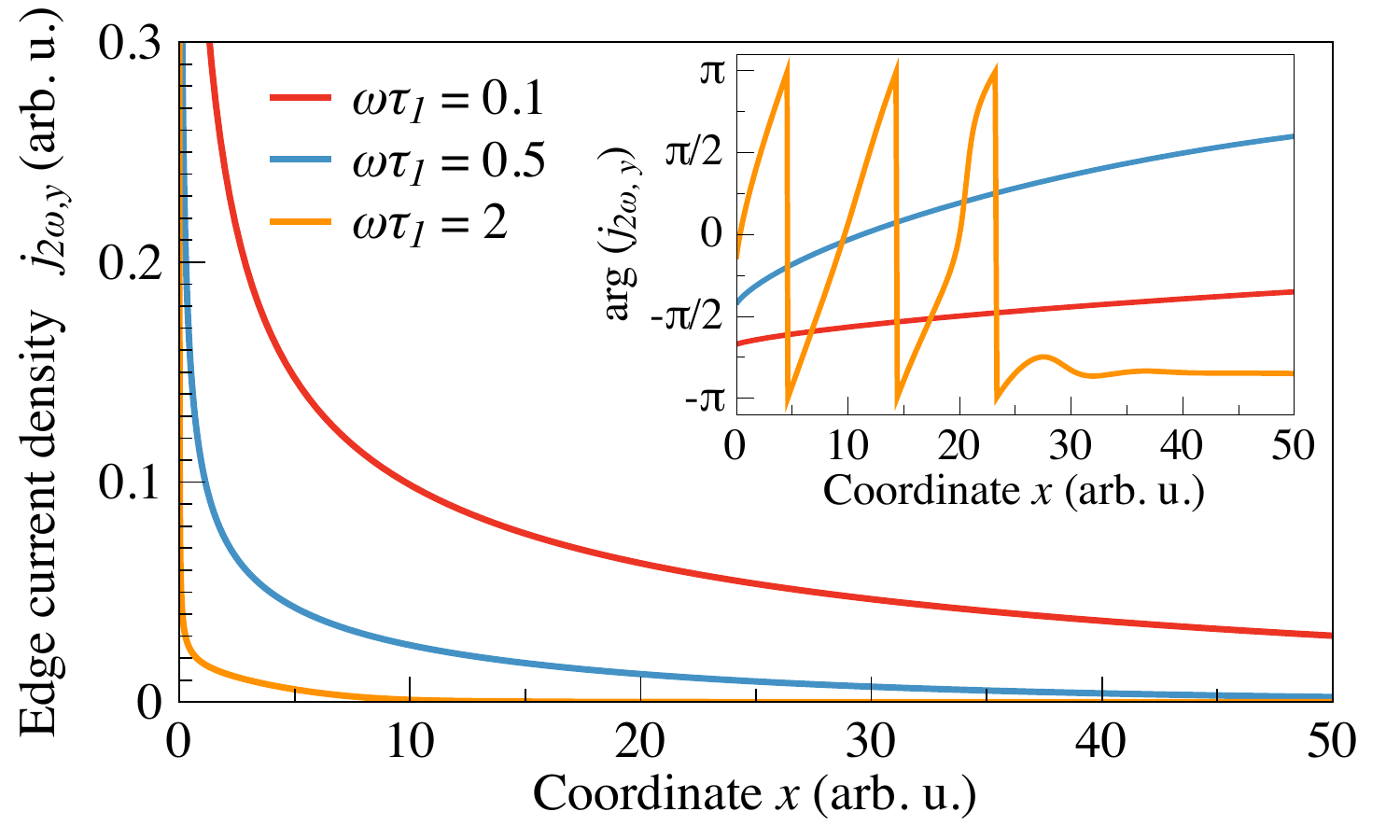}
\caption{\label{fig3} Spatial profile of the edge current density $j_{2\omega,y} (x)$ induced by linearly polarized field $E_{\omega, x} = E_{\omega, y}$. 
The main graph and the inset show the modulus $|j_{2\omega, y}|$ and the argument $\mathrm{arg}(j_{2\omega, y})$ of the complex-value current $j_{2\omega, y}$, respectively. The current density is calculated numerically in the local response approximation.
}
\end{figure}

Figure~\ref{fig3} shows the spatial distributions of the current density $j_{2\omega,y} (x)$ near the edge. 
Different curves correspond to different $\omega\tau_1$. 
The distributions are obtained from Eq.~\eqref{jy} for $\tau_2 \ll \tau_1$, when the first term in Eq.~\eqref{jy} is negligible. The second term is then found by numerical calculations of the charge density $\rho_\omega(x)=e\nu \sum_{\bm p} f_1$ in the local response approximation (see next section for details). In this approximation, the decay of the edge current $j_{2\omega,y}(x)$ in the 2D bulk 
is determined by the length of dynamical screening $l_{\rm scr} = \sigma_0/\omega$. Hence, the current profile narrows with the frequency increase.  At large $\omega \tau_1$, $\mathrm{arg}(j_{2\omega, y})$ exhibits spatial oscillations, i.e., the currents $j_{2\omega, y}$ at different $x$ are phase-shifted and flow in the opposite directions in the nearby regions. These oscillations are caused by excitation of the edge plasmons, see App.~\ref{LRA_app} for details.

\section{Current normal to the edge}
 
Now consider the $x$ component of the edge current. Multiplying Eq.~\eqref{f2} by $v_x$, summing up over $\bm p$, and taking into account 
that $\sum_{\bm p} v_x \partial f_1/\partial p_y =0$ and $\sum_{\bm p} v_x \partial f_n /\partial p_x = - (1/m) \sum_{\bm p} f_n$, we obtain
\begin{equation}
\label{jx}
j_{2\omega, x} = \frac{-e \nu \tau_1}{1 - 2\i\omega \tau_1} \left[ \sum \limits_{\bm p} v_x^2 \pderiv{f_2}{x} - \frac{e \mathcal E_{\omega, x}}{m} \sum \limits_{\bm p} f_1 \right]  + \sigma_{2\omega} \mathcal E_{2\omega,x} \:,
\end{equation} 
where $\sigma_{2\omega} = \sigma_0/(1 - 2 \i \omega \tau_1)$ is the conductivity at double frequency.

The total current is obtained from Eq.~\eqref{jx} by integrating over $x$. Using the relation $\sum_{\bm p} f_1 = -(\i/\omega) \sum_{\bm p} v_x \partial f_1/\partial x$, we obtain
\begin{multline}
\label{Jx}
J_{2\omega, x} = -\frac{e \nu \tau_1}{1 - 2\i\omega \tau_1} \sum \limits_{\bm p} v_x^2 \left[ f_2(\bm p, +\infty) - f_2(\bm p, 0) \right]  \\ -\frac{\i e^2 \nu \tau_1 E_{\omega, x}}{m \omega(1 - 2\i\omega \tau_1)} \sum \limits_{\bm p} v_x [f_1(\bm p, +\infty) - f_1(\bm p, 0)] \\
 + \frac{\i e^2 \nu \tau_1}{m \omega(1 - 2\i\omega \tau_1)}  \int  \frac{d{\cal E}_{\omega,x}}{dx} \sum \limits_{\bm p} \ v_x f_1 \, dx
+ \sigma_{2\omega} \int \mathcal E_{2\omega, x}(x) dx \:.
\end{multline}
Comparing Eqs.~\eqref{Jy} and \eqref{Jx} one observes that the current perpendicular to the edge contains two additional contributions which are not proportional to a difference between the distribution function at the edge and in the 2D bulk. Evaluation of these terms requires the knowledge of the distribution function corrections $f_1$ and $f_2$ and the electric field $\mathcal E_x$ in the whole half-space $x > 0$. The corrections and the field can be found numerically from Eqs.~\eqref{f1}-\eqref{screen2}. However, solving Eqs.~\eqref{f1}-\eqref{screen2} self-consistently is, in general, a challenging task.
Therefore, in what follows we consider two approximations.

Note that symmetry consideration of the edge SHG allows the current $J_{2\omega, x}$ to be induced by $E_{\omega, x}^2$ and $E_{\omega, y}^2$. However, the analysis of Eqs.~\eqref{jx},~\eqref{f1}, and~\eqref{f2} shows that, for specular reflection of electrons from the edge, $J_{2\omega, x} \propto E_{\omega, x}^2$, i.e., the current $J_{2\omega, x}$ vanishes for the field $\bm E_{\omega}$ polarized along the edge. This result also holds in the local response approximation considered below.


\subsection{Strong screening. Local response approximation}

In the absence of high-$\varepsilon$ dielectric environment, 
Coulomb interaction in 2D systems is dominant and, therefore, 
drift currents induced by local electric fields prevail over diffusion currents. In the local response approximation~\cite{Mikhailov2005}, 
terms with the spatial gradients in equations for the current density are neglected. As a result, equations for the current density 
at $\omega$ and $2\omega$ assume the form
\begin{equation}
\label{jw_local}
j_{\omega, x}(x) = \sigma_\omega \mathcal E_{\omega,x} (x) \:
\end{equation}
and
\begin{equation}
\label{j2w_local}
j_{2\omega, x}(x) = \frac{e \tau_1 \rho_{\omega}(x) \mathcal E_{\omega, x}(x)}{m(1 - 2\i\omega \tau_1)}   
+ \sigma_{2\omega} \mathcal E_{2\omega,x}(x) \:.
\end{equation} 
The latter follows directly from Eq.~\eqref{jx}. 


To find the spatial profile of the current density $j_{2\omega, x}(x)$ and the total current $J_{2\omega,x}$ we solve Eqs.~\eqref{jw_local} and
~\eqref{j2w_local} self-consistently with Eqs.~\eqref{screen2} for $\mathcal E_{n\omega,x}(x)$  and the continuity equations $- \i n \omega \rho_{n \omega} + d j_{n \omega,x}/dx = 0$, see Appendix~\ref{LRA_app} for details. The absence of the current through the edge implies the boundary conditions $j_{n\omega, x}(0) = 0$.


\begin{figure}[htpb]
\includegraphics[width=0.49\textwidth]{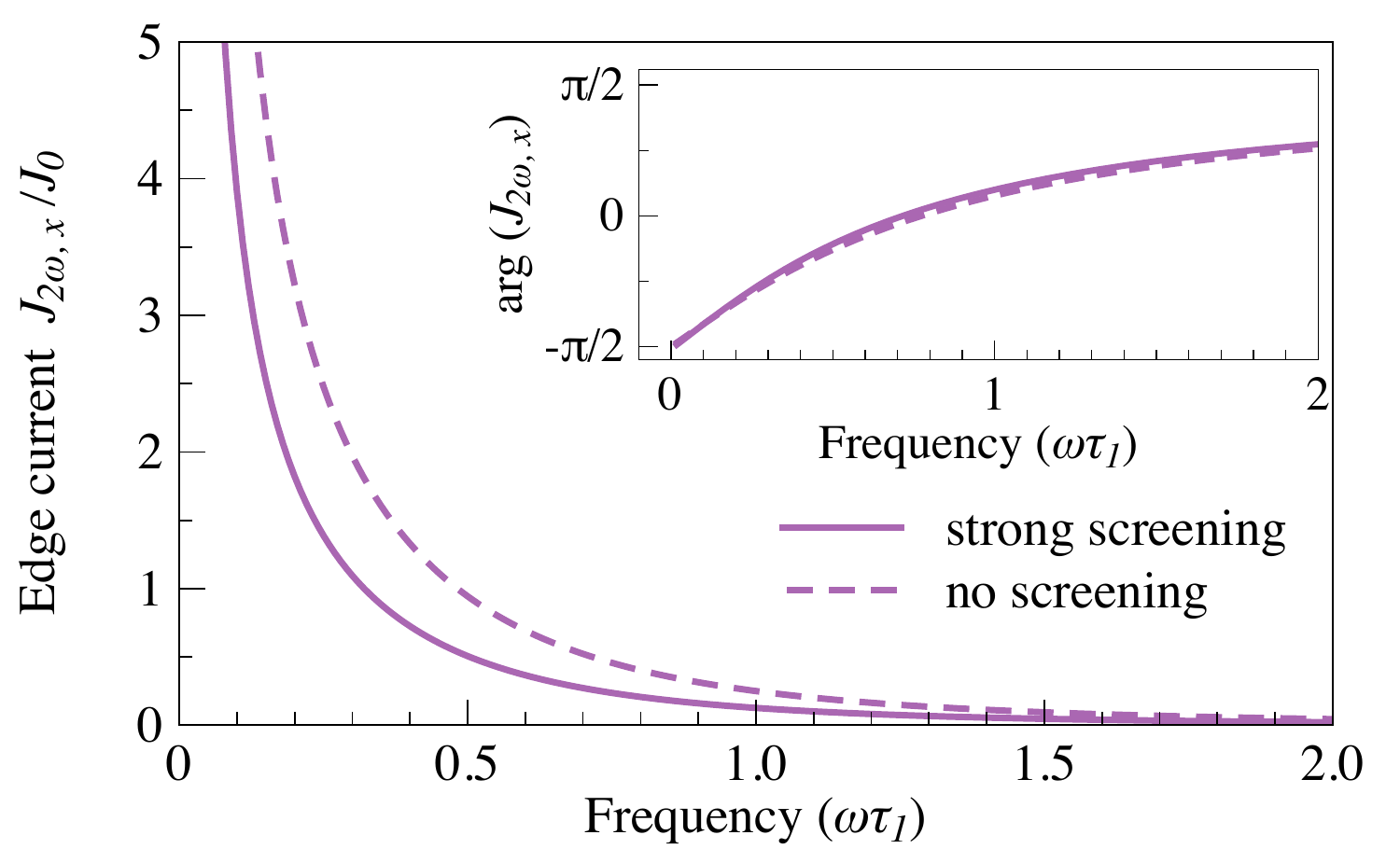}
\caption{\label{fig4} Frequency dependence of the edge current at $2\omega$ flowing perpendicular to the edge.  
The solid line corresponds to the local response approximation in the regime of strong screening. The dashed line is calculated 
in the hydrodynamic regime with $\tau_2 \ll \tau_1$ and neglecting screening. 
 The main graph and the inset show the modulus $|J_{2\omega, x}|$ and the argument $\mathrm{arg}(J_{2\omega, x})$ of the complex-value current $J_{2\omega, x}$, respectively. The current is measured in $J_0 = e \sigma_0 \tau_1^2 E_{\omega,x}^2/m$.}
\end{figure}

\begin{figure}[htpb]
\includegraphics[width=0.49\textwidth]{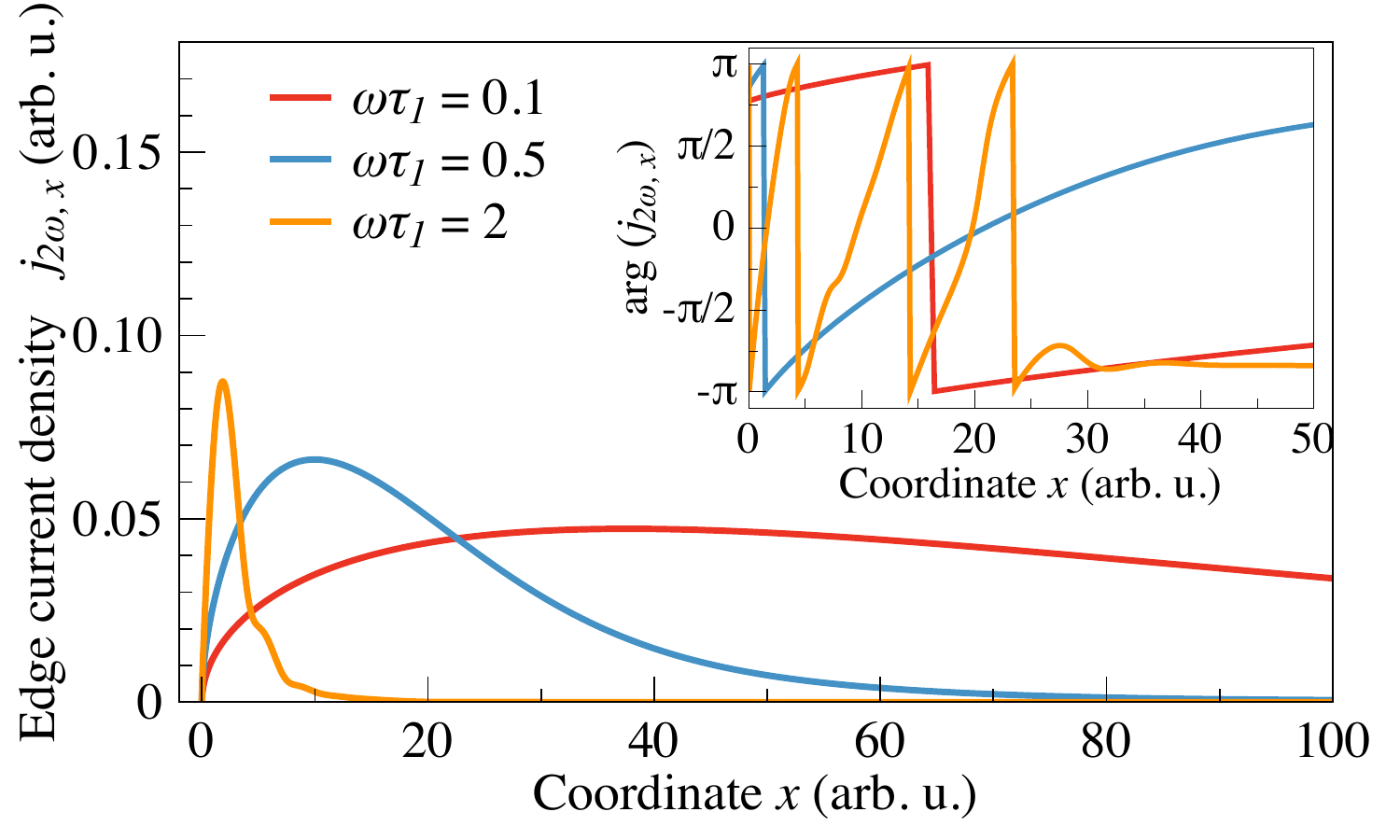}
\caption{\label{fig5} Spatial profile of the edge current density $j_{2\omega,x} (x)$ induced by linearly polarized field $\bm E_{\omega} \parallel x$. 
The main graph and the inset show the modulus $|j_{2\omega, x}|$ and the argument $\mathrm{arg}(j_{2\omega, x})$ of the complex-value current $j_{2\omega, x}$, respectively. The current density is calculated numerically in the local response approximation, Eq.~\eqref{j2w_local}.
}
\end{figure}

Figure~\ref{fig4} shows the frequency dependence of the current $J_{2\omega, x}$. The solid line shows the current calculated numerically in the local response approximation for linearly polarized incident field $\bm E_{\omega} \parallel x$. The dependence closely follows the one for $J_{2\omega, y}$ shown in Fig.~\ref{fig2}, and the phase shift between $J_{2\omega, x}$ and $J_{2\omega, y}$ for linearly polarized incident field is close to zero.

Figure~\ref{fig5} shows the spatial distributions of the current density $j_{2\omega,x} (x)$ near the edge. 
Different curves correspond to different $\omega\tau_1$. Similarly to the current along the edge $j_{2\omega,y}$, 
the current $j_{2\omega,x}$ decays in the 2D bulk on the scale of the screening length $l_{\rm scr} = \sigma_0/\omega$ and its profile narrows with the frequency increase. In contrast to $j_{2\omega,y}$, the current $j_{2\omega,x}$ vanishes at $x = 0$ as set by boundary conditions.
Similarly to $j_{2\omega,y}$, the profile of $j_{2\omega, x}$ exhibits spatial oscillations at large $\omega \tau_1$ caused by the excitation of edge plasmons.

\subsection{Negligible screening} 

The opposite case of weak screening is realized if the 2D layer is surrounded by a high-$\eps$ dielectric medium and one can neglect the back action of an in-plane electric field produced by charge oscillations. In this case, the total electric field $\bm{\mathcal E}$ acting upon the electrons coincides with the incident field $\bm E_{\omega}$ and the last line in Eq.~\eqref{Jx} vanishes. To calculate the other contributions to $J_{2\omega,x}$
we need to find the difference between the distribution functions at the edge and in the bulk by solving Eqs.~\eqref{f1} and~\eqref{f2} with ${\mathcal E}_{\omega, x} = E_{\omega}$, ${\mathcal E}_{\omega, y} = 0$, and $\bm{\mathcal E}_{2\omega} = 0$.  We do it analytically for $\tau_2 \ll \tau_1$, which corresponds to the hydrodynamic regime of electron flow, and $\omega \tau_2 \ll 1$. 
In this regime, one can retain only the zeroth and first angular harmonics in the distribution function corrections $f_1$ and $f_2$.

The functions $f_n(\bm p,x)$ ($n=1,2$) can be searched in the form $f_n(\bm p, x) = a_n(p, x) + v_x b_n(p, x)$.
The absence of the current through the edge, the current at $2\omega$ in the bulk, and the charge/energy oscillations at $\omega$ in the bulk
implies $b_n(p, 0) = 0$, $b_2(p, +\infty) = 0$, and $a_1(p,+\infty) = 0$, respectively.
Solution of Eq.~\eqref{f1} with these boundary conditions has the form  
\begin{equation}
a_1 = e \lambda_\omega^{-1} E_{\omega, x} f_0' \e^{- \lambda_\omega x} ,\;
b_1 = -e \tau_{\omega} E_{\omega, x} f_0'  \left( 1 - \e^{- \lambda_\omega x} \right) ,
\end{equation}
where 
\begin{equation}
\label{lambdaw}
\tau_{\omega} = \frac{\tau_1}{1 - \i\omega\tau_1} \:,\;\;
\lambda_\omega = (1-\i) \sqrt{\frac{m \omega}{2 \eps \tau_\omega}}\:,
\end{equation}
$\eps = p^2/2m$, and $(...)' = \partial (...)/ \partial \eps$. Equation~\eqref{f2} leads to the system of coupled differential equations
\begin{eqnarray}
\label{b0b1}
-2\i m\omega a_2 +\eps \pderiv{b_2}{x} &=& - e E_{\omega,x} (b_1 \eps)' , \nonumber \\
\pderiv{a_2}{x} + (\tau_1^{-1} - 2\i\omega) b_2 &=& - e E_{\omega, x} a_1' \:.
\end{eqnarray}
Its solution has the form 
\begin{eqnarray}
a_2 &=& A + B \e^{-\lambda_\omega x} + C x \e^{-\lambda_\omega x} - D \e^{-\lambda_{2\omega} x}/ (\lambda_{2 \omega} \tau_{2\omega})\:, \nonumber \\
b_2 &=& D \left( \e^{-\lambda_\omega x} - \e^{-\lambda_{2\omega} x} \right) + F x \e^{-\lambda_\omega x} \:, 
\end{eqnarray}
where $\tau_{2\omega}$ and $\lambda_{2\omega}$ are given by Eq.~\eqref{lambdaw} with $\omega \to 2\omega$, and the constants 
$A$, $B$, $C$, $D$, $F$ are found from Eq.~\eqref{b0b1}.

Further, we note that   
\[ 
e\nu \sum_{\bm p} v_x f_1(\bm p, +\infty)  = \sigma_\omega E_{\omega,x} \:,~  \sum_{\bm p} v_x f_1(\bm p, 0) = 0 \:,
\] 
and 
\[
\sum_{\bm p} v_x^2 [f_2(\bm p, +\infty) - f_2(\bm p, 0)] = \frac{1}{2m} \sum_{\bm p} \eps [a_2(p, +\infty) - a_2(p, 0)] \,.
\]
By solving Eq.~\eqref{b0b1} we find 
\begin{multline}
\sum_{\bm p} \eps [a_2(p, +\infty) - a_2(p, 0)]  = \frac{\i \sigma_\omega E_{\omega,x}^2}{\nu \, \omega }
\\ \times \frac{2 \tau_\omega^2 + 4 \tau_{2\omega} \tau_\omega - \tau_{2\omega}^2 + (3/4) \sqrt{2\tau_{\omega}} \sqrt{\tau_{2\omega}} (\tau_{2\omega} - 6 \tau_\omega)}{(\tau_{2\omega} - 2\tau_\omega)^2}\:,
\end{multline}
where the sign of $\sqrt{\tau_{n \omega}}$ is chosen so that $\mathrm{Re} \sqrt{\tau_{n \omega}} > 0$. 

Finally, Eq.~\eqref{Jx} for the current $J_{2\omega, x}$ yields  
\begin{equation}
\label{J2wx_hydro}
J_{2\omega, x} = -\frac{\i e \sigma_0 \tau_1  E_{\omega, x}^2}{m \omega (1 - \i\omega \tau_1)(1 - 2\i\omega \tau_1)} F(\omega)\:,
\end{equation}
where
\begin{equation}
F(\omega) = \frac{3[8 \tau_\omega^2 + \sqrt{2\tau_\omega} \sqrt{\tau_{2\omega}}(\tau_{2\omega} - 6\tau_\omega)]}{4(\tau_{2\omega} - 2\tau_\omega)^2}\:.
\end{equation}
The frequency dependence of the current $J_{2\omega, x}$ given by Eq.~\eqref{J2wx_hydro} is shown in Fig.~\ref{fig4} by the dashed line. 
Comparing the dashed and solid lines we see that the currents at $2\omega$ calculated in the regimes of strong and negligible screening are close in magnitude.

We recall that Eq.~\eqref{J2wx_hydro} is obtained in the hydrodynamic regime with $\tau_2 \ll \tau_1 , \omega^{-1}$. Within the same approximation, the current along the edge $J_{2\omega,y}$ is given by Eq.~\eqref{J2wy_final} with $\tau_2 = 0$. In this regime, the ratio $J_{2\omega,x}/J_{2\omega,y}$ is determined by the incident field polarization and the function $F(\omega)$. The absolute value of $F(\omega)$ lies in the range $0.7 - 0.8$ and its argument is close to zero in the whole frequency range.  The latter suggests that the ac current $\bm {J}_{2\omega}$ is linearly polarized for linearly polarized field $\bm E_{\omega}$.

\section{Summary} \label{summary}

To summarize, we have studied theoretically second harmonic generation (SHG) emerging at the edge of a two-dimensional (2D) electron gas. 
It has been shown that ac in-plane electric field oscillating at frequency $\omega$ induces ac electric current at frequency $2 \omega$ near the edge.
The current is formed in the edge region determined by the dynamical screening of the electric field and the mean free path of electrons. The edge current $\bm J_{2\omega}$ has both parallel and perpendicular to the edge components, 
$J_{2\omega, \parallel} \propto E_{\omega,\parallel} E_{\omega,\perp}$ and $J_{2\omega, \perp} \propto E_{\omega, \perp}^2$, respectively, 
where $\bm E_{\omega}$ is the driving electric field amplitude. The currents $J_{2\omega, \parallel}$ and $J_{2\omega, \perp}$ emit 
electromagnetic fields at $2\omega$ with different polarizations and different radiation patterns.
We have developed the kinetic theory of high-frequency nonlinear edge transport which takes into account the screening of the in-plane electric field 
by 2D electrons. The parallel current $J_{2\omega, \parallel}$ is calculated analytically in a quite general case whereas the perpendicular current $J_{2\omega, \perp}$ is calculated in the limiting cases of strong and negligible screening. At $\omega \tau_1 > 1$, where $\tau_1$ is the momentum relaxation time, the spatial profile of the current density contains oscillations caused by the excitation of edge plasmons.
The SHG spectroscopy can be used to visualize edges and spatial inhomogeneities in doped 2D materials and heterostructures.

\acknowledgements 

M.V.D. acknowledges financial support from the Russian Science Foundation (Project No. 21-72-00047) and the Basis Foundation for the Advancement of Theoretical Physics and Mathematics. 

\appendix

\section{Profiles of electric field, charge, and current in the local response approximation} \label{LRA_app}

Here, we calculate the spatial distributions of the electric field, charge, and electric current at the edge of a 2D electron gas in the local response approximation. To this end, we 
solve Eqs.~\eqref{screen2}, \eqref{jw_local}, and~\eqref{j2w_local} together with the continuity equations 
\begin{equation}\label{rhow1}
- \i n \omega \rho_{n \omega} + \frac{d j_{n \omega,x}}{d x} = 0
\end{equation}
inside a strip of the width $2a$ occupying $-a \leq x \leq a$.

Equation~\eqref{screen2} for the strip assumes the form
\begin{equation}
\label{Ew_strip}
\mathcal E_{n\omega, x}(x) = E_{\omega, x} \delta_{n,1} + \int\limits_{-a}^{a}\frac{2 \rho_{n\omega}(x') dx'}{x-x'} \:.
\end{equation}
This integral equation can be inverted to express $\rho_{n\omega}$ via $\mathcal E_{n\omega, x}$. Taking into account that $\rho_{n\omega}$ is infinite at $x = \pm a$ and $\int_{-a}^{a} \rho_{n\omega}(x) dx = 0$, we obtain~\cite{Polyanin_book}
\begin{multline}
\label{rhow2} 
\rho_{n\omega}(x) =  \frac{x E_{\omega,x} \delta_{n,1} }{2\pi \sqrt{a^2-x^2}} 
\\ - \frac{1}{2\pi^2} \int \limits_{-a}^{a} \frac{\sqrt{a^2-x'^2} \mathcal E_{n\omega,x}(x')dx'}{\sqrt{a^2-x^2} (x-x')}  \:.
\end{multline}
Note that the first term in the right-hand side of Eq.~\eqref{rhow2} at $n=1$ gives the distribution of charge density induced by the static electric field, i.e., in the limit  $\omega \rightarrow 0$, when the $x$ component of the field inside the strip is completely screened so that $\mathcal E_{\omega, x} = 0$. 

Equations~\eqref{rhow1} and~\eqref{rhow2} yield 
\begin{multline}
\label{field} 
\frac{-\i}{n \omega} \frac{d j_{n\omega,x}}{d x} =  \frac{x E_{\omega,x} \delta_{n,1} }{2\pi \sqrt{a^2-x^2}} 
\\ - \frac{1}{2\pi^2} \int \limits_{-a}^{a} \frac{\sqrt{a^2-x'^2} \mathcal E_{n\omega,x}(x')dx'}{\sqrt{a^2-x^2} (x-x')}  \:.
\end{multline}

First, we analyze the linear response and calculate the distributions $\mathcal E_{\omega, x}(x)$, $\rho_{\omega}(x)$, and $j_{\omega, x}(x)$.
By integrating Eq.~\eqref{field} from $-a$ to $x$ and using Eq.~\eqref{jw_local} together with the boundary condition $j_{\omega, x}(-a) = 0$
we obtain
\begin{multline}
\label{int_Ew1}
\frac{\i l_{\rm scr} \mathcal E_{\omega, x}}{1 - \i\omega \tau_1}  = \frac{\sqrt{a^2-x^2} E_{\omega,x}}{2\pi}  \\
+  \int \limits_{-a}^{x} \frac{d x''}{2\pi^2} \int \limits_{-a}^{a} \frac{\sqrt{a^2-x'^2} \mathcal E_{\omega,x}(x')dx'}{\sqrt{a^2-x''^2} (x''-x')} \:.
\end{multline} 
Equation~\eqref{int_Ew1} is simplified further by introducing the variables $\alpha$ and $\beta$ as follows: $x = a \cos \alpha$ and $x' = a \cos \beta$. 
Taking into account that 
\begin{equation}
\int \limits_{-a}^x \frac{d x''}{\sqrt{a^2-x''^2} (x'-x'')} = \frac{1}{a \sin \beta} \ln \left( \frac{\sin \frac{\alpha + \beta}{2}}{|\sin \frac{\alpha - \beta}{2}|} \right) ,
\end{equation}
we finally obtain the integral equation for $\mathcal E_{\omega, x}$ 
\begin{multline} 
\label{int_Ew2}
\frac{\i l_{\rm scr} \mathcal E_{\omega, x}(\alpha)}{a (1 - \i \omega \tau_1)}  = \frac{E_{\omega, x} \sin \alpha}{2\pi}  \\
- \frac{1}{2\pi^2} \int \limits_0^\pi \ln \left( \frac{\sin \frac{\alpha + \theta}{2}}{|\sin \frac{\alpha - \beta}{2}|} \right) \sin \beta~\mathcal E_{\omega, x} ( \beta) d\beta \:.
\end{multline}

Equation~\eqref{int_Ew2} can be solved by decomposing the electric field $\mathcal E_{\omega, x}(\alpha)$ in the Fourier series
\begin{equation}
\label{Ew_series}
\mathcal E_{\omega, x}(\alpha) = \sum \limits_{m=1}  \mathcal E_{\omega}^{(m)} \sin m\alpha\:.
\end{equation}
The integrals in Eq.~\eqref{int_Ew2} are calculated analytically,  
\begin{equation}
\label{int_ln}
\int \limits_0^\pi \ln \left( \frac{\sin \frac{\alpha + \beta}{2}}{|\sin \frac{\alpha - \beta}{2}|} \right) \sin m \alpha~d\alpha = \frac{\pi}{m} \sin m\beta\:.
\end{equation}
This procedure allows us to reduce the integral Eq.~\eqref{int_Ew2} to the set of linear equations 
\begin{equation}
\label{Ew_matrix}
\frac{\i l_{\rm scr} \pi \mathcal E_{\omega}^{(m)}}{a(1 - \i\omega \tau_1)} + \frac{1}{\pi} \sum \limits_{m'=1}^{\infty} \mathcal K_{mm'} \mathcal E_{\omega}^{(m')} = \frac{E_{\omega, x}}{2} \delta_{m,1}\:,
\end{equation}
where
\begin{equation}
\mathcal K_{mm'} = 
-\dfrac{2 [1 + (-1)^{m+m'}] m'}{m'^4 + (m^2-1)^2 - 2m'^2(m^2+1)}\:. 
\end{equation} 
The Fourier components $\mathcal E_{\omega}^{(m)}$ can be readily found from the numerical solution of the equation set~\eqref{Ew_matrix}.

\begin{figure}[htpb]
\includegraphics[width=0.49\textwidth]{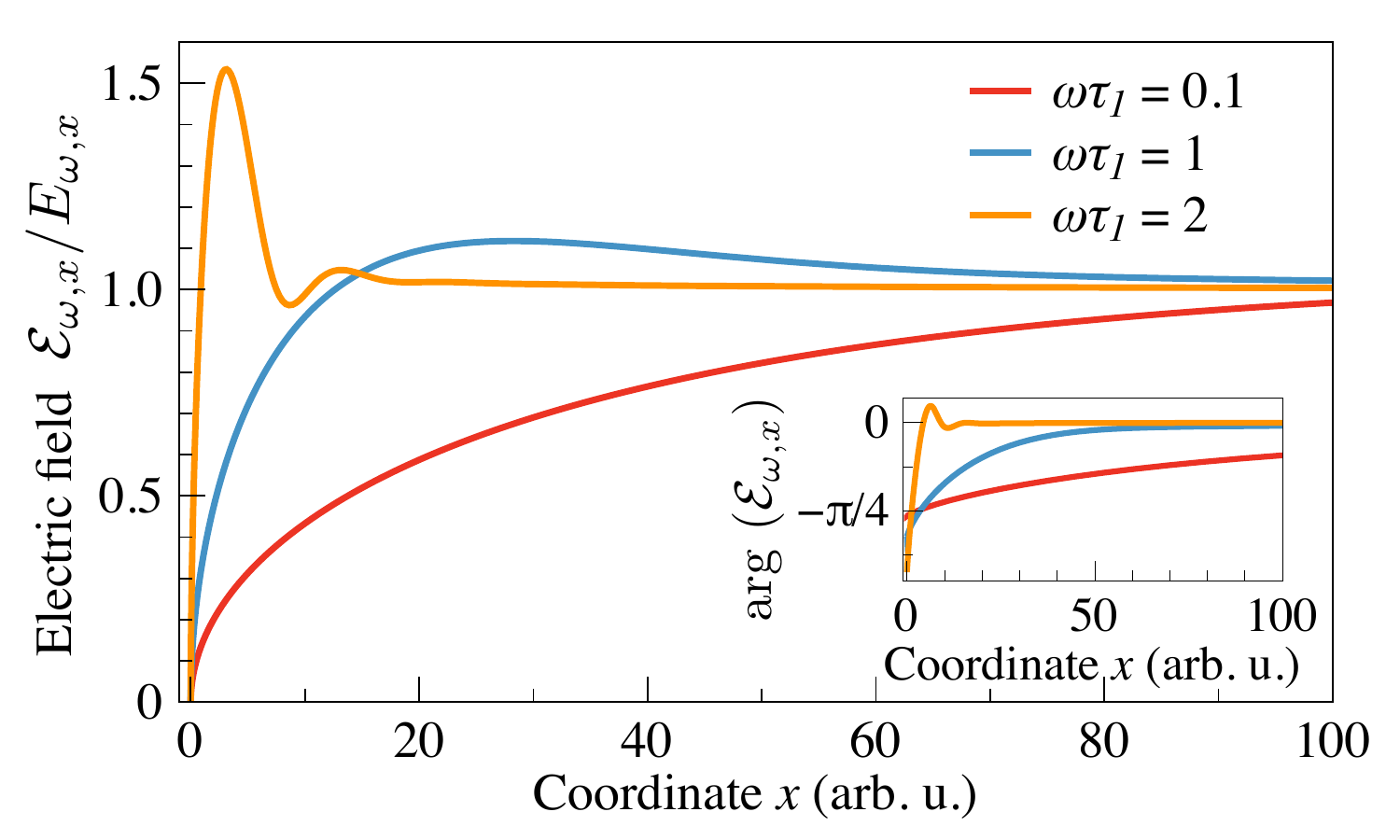}
\caption{\label{figA1} Spatial profiles of the electric field $\mathcal E_{\omega,x} (x)$ at the edge of 2D electron gas subject to the incident electric field $E_{\omega,x}$. The main graph and the inset show the modulus $|\mathcal E_{\omega, x}|$ and the argument $\mathrm{arg}(\mathcal E_{\omega, x})$ of the complex-value field $\mathcal E_{\omega, x}$, respectively. The field profile is calculated numerically in the local response approximation.}
\end{figure}

Figure~\ref{figA1} shows the spatial profiles of the electric field $\mathcal E_{\omega, x} (x)$ at the edge of 2D electron gas calculated after Eqs.~\eqref{Ew_series}, \eqref{Ew_matrix} for different frequencies of the incident field $E_{\omega, x}$. The $x$ coordinate in Fig.~\ref{figA1} is counted from the left edge of the wide strip. The calculations are done for the strip width $a \gg l_{\rm scr}$ when field profiles at the edge do not depend on $a$. 
Figure~\ref{figA1} reveals that the electric field is efficiently screened in the region $x \sim l_{\rm scr}=\sigma_0 /\omega$ near the edge 
($\mathcal E_{\omega, x} \sim \sqrt{x}$ at small $x$), whereas far from the edge the field is unscreened and coincides with $E_{\omega, x}$.
With the frequency increase, the screening length decreases and the region of field screening narrows. At large $\omega \tau_1$, the profile 
$\mathcal E_{\omega, x}(x)$ contains oscillations caused by excitation of plasmons with the wave vectors $q \sim \omega^2/(2\pi n_e e^2) = \omega \tau_1/(2\pi l_{\rm scr})$~\cite{Mikhailov2005, Zabolotnykh2019}. 

Now we calculate $j_{2\omega, x}(x)$. 
By integrating Eq.~\eqref{field} with $n=2$ over $x$ from $-a$ to $x$ and using the boundary condition $j_{2\omega, x}(-a) = 0$ and 
Eq.~\eqref{j2w_local}, one obtains
\begin{multline}
\label{int_j2w}
-\frac{\i l_{\rm scr} j_{2\omega, x}(\alpha)}{2a(1 - 2\i \omega \tau_1)}  =  \frac{1}{2\pi^2} \int \limits_0^\pi \ln \left( \frac{\sin \frac{\alpha + \beta}{2}}{|\sin \frac{\alpha - \beta}{2}|} \right) \sin \beta \\
\times \left[ j_{2\omega, x}(\beta) -  \Lambda  \mathcal E_{\omega, x} (\beta) \frac{d \mathcal E_{\omega, x}}{dx} (\beta) \right] d\beta\:,
\end{multline}
where
\begin{equation}
\Lambda= -\frac{\i e \sigma_0 \tau_1}{m \omega (1 - \i\omega \tau_1) (1 - 2\i\omega \tau_1)}
\end{equation} 
and $\mathcal E_{\omega, x}$ is the field at $\omega$ calculated above.

Decomposing the current $j_{2\omega, x}(\alpha)$ in the Fourier series,
\begin{equation}
\label{j2w_series}
j_{2\omega, x}( \alpha) = \sum \limits_{m=1}  j_{2\omega}^{(m)} \sin m\alpha \:,
\end{equation}
and using Eqs.~\eqref{Ew_series} and~\eqref{int_ln},
we reduce the integral Eq.~\eqref{int_j2w} to the set of linear equations
\begin{multline}
\label{j2w_matrix}
\frac{\i l_{\rm scr} \pi j_{2\omega}^{(m)}}{2 a(1-2\i\omega \tau_1)} + \frac{1}{\pi^2} \sum \limits_{m'=1}^{\infty} \mathcal K_{mm'} j_{2\omega}^{(m')} =
  \\ =  -\frac{\Lambda }{4a} \left[ \sum_{m'=1}^{m-1} \frac{m'}{m}  \mathcal E_{\omega}^{(m')} \mathcal E_{\omega}^{(m-m')} - \sum_{m'=m+1}^{\infty} \mathcal E_{\omega}^{(m')} \mathcal E_{\omega}^{(m'-m)} \right] \:.
\end{multline}
The total current flowing near the left edge is found as 
\begin{equation}
\label{J2w_local}
J_{2\omega,x} = \int_{-a}^0 j_{2\omega,x} dx = 
a \sum_{m>1} \frac{m \cos (\pi m/2)}{m^2 - 1} j_{2\omega}^{(m)} \:.
\end{equation}

The profile of the edge current calculated after Eqs.~\eqref{j2w_series}-\eqref{j2w_matrix} is shown in Fig.~\ref{fig5} of the main text. 
The total current $J_{2\omega, x}$ is shown in Fig.~\ref{fig4} by the solid line.

\end{document}